
\magnification\magstep 1
\vsize=22 true cm
\hsize=16 true cm
\baselineskip=0.9 true cm
\parindent=1.1 true cm

\raggedbottom

\def\intif{\int_0^\infty}

\font\bfmag=cmbx10 scaled\magstep3
\font\bfmagd=cmbx10 scaled\magstep2
\def\tr{{\rm Tr}}
\def\Lag{{\cal L}_{\pi\rho}}
\def\WMN{{W_{\mu\nu}}}

\def\L6{{\cal L}_{\omega}}

\rightline {}
\rightline {}
\bigskip
\centerline {\bfmag On the role of vector mesons}
\centerline {\bfmag in topological soliton stability}
\bigskip
\centerline{\bf Abdellatif Abada, Dimitri Kalafatis$^*$ and Bachir Moussallam}
\smallskip
\centerline{\sl Division de Physique Th\'eorique\footnote\dag{
Unit\'e  de Recherche des Universit\'es Paris XI et Paris VI
associ\'ees au CNRS. }}
\centerline{\sl Institut de Physique Nucl\'eaire, 91406 Orsay Cedex}
\centerline{\sl $^*$ and LPTPE, Universit\'e Pierre et Marie
Curie,}
\centerline{\sl 4 Place Jussieu 75252 Paris Cedex 05 (France)}
\bigskip
\bigskip
\bigskip
\centerline{\bfmagd Abstract}
Isospin one vector mesons (in particular the $\rho$) are usually
described as massive Yang-Mills particles
in the chiral Lagrangian. We investigate some aspects of an alternative
approach in the soliton sector. It is found that the soliton
is stable in very much the same way as with the $\omega$-meson and that
spontaneous parity violating classical solutions do not exist. The
formulation in terms of antisymmetric tensors is shown to be canonically
related to a vector field description provided the Skyrme term is added
to the latter.

\vfill
\noindent IPNO/TH 92-83 \hfill {October 1992}

\eject
\rightline{}
\noindent{\bfmagd I. Introduction}

During the last ten years, a great deal of effort has been devoted to
the study of baryons as topological solitons of nonlinear meson theories.
This possibility emerged a long time ago in the pioneering work of Skyrme [1].
Later, the connection between solitons of meson theories, and
large $N_c$ expansion of QCD [2] was established [3]. Low-energy baryon
phenomenology was considered in Ref. [4]. In this model the  effective
Lagrangian is based on the nonlinear
$\sigma$ model and an antisymmetric term of fourth order in powers of
the derivatives of the pion field (the so-called
Skyrme term) is added in order to avoid soliton
collapse [5].

It was soon realized [6], that the Skyrme term can be thought as a large
mass (local) limit of a model with $\rho$-mesons. Similarly $\omega$-meson
exchange generates a term of order six. These findings provided the
ground for the construction of a more
general effective Lagrangian, which includes the low lying mesons
and considerably improves the predictions for baryon physics [7].
 An important question is that of soliton
stability in the presence of vector mesons. The $\omega$ field, was
shown to yield stable
solutions in Ref. [8]. But this was not the case of the low-lying
isovector-vector mesons (and the $\rho$ in particular), which were
introduced as hidden gauge particles [9] as well as massive Yang-Mills
[10] fields, and in both cases the soliton was shown to be unstable ,
contrary to earlier claims [11]. This different behaviour of the $\rho$
and the $\omega$-mesons may look surprising as their local large mass
limits are both stabilizing the soliton.

Now, in another context, the role of meson resonances in chiral perturbation
theory was investigated by Ecker et al. [12]. In that paper, isovector
mesons are described in terms of antisymmetric tensor fields (first
proposed in Ref. [13]). They are introduced in the chiral Lagrangian
following the standard prescription of nonlinear chiral symmetry for
massive particles [14]. It was
shown in Ref. [15] that this approach is equivalent to the hidden
symmetry (or Yang-Mills) one to $O(p^4)$ in the
chiral expansion.

The purpose of this letter is to examine whether the introduction of the
$\rho$-meson in this alternative approach could play a different
role {\it in the baryon sector}. We will focus our attention then to soliton
stability in such a model. Particular attention will be given to the
similar role that the $\rho$ and the $\omega$ fields
are finally playing in soliton stability.

The Lagrangian is introduced in section II. Soliton solutions of
winding number one are then constructed in section III. We study
the classical
stability of these solutions in section IV. We finally
discuss some issues on the vector field approach in section V.

\noindent{\bfmagd II. The model}

Compared to the conventional way of introducing isospin-one vector
resonances in the chiral
expansion, the method of Refs. [12,13] has one basic difference:
the assumption that the resonance fields {\it are not} gauge bosons of
any kind. They are simply assumed to transform as $R\to
h(\pi)Rh^\dagger(\pi)$ where $h$ is defined from the transformation
property of $u=U^{1/2}$ : $u\to g_Luh^\dagger=hug_R^\dagger$ such that
$U=\exp(i\vec\tau.\vec\pi/f_{\pi})$ transforms linearly. It proves
convenient ( see sec. V.) then to describe spin one mesons in terms of
antisymmetric tensor fields $W_{\mu\nu}$. The following Lagrangian
ensures that only three physical degrees of freedom actually propagate
[12] and is the simplest one compatible with chiral symmetry:

$$\eqalign{
\Lag=&-{1\over
2}\tr(\nabla^{\mu}\WMN\nabla_{\sigma}W^{\sigma\nu})+{{M_{\rho}^2}\over 4}
\tr(\WMN
W^{\mu\nu})\cr
&+{{f_\pi^2}\over 4}\tr(\partial_\mu U\partial^\mu U^\dagger)
+i{{G_\rho}\over{2\sqrt{2}}}\tr(\WMN [u^\mu, u^\nu])\cr}\eqno(1)
$$
\noindent where,
$$\eqalign{
&\nabla_\mu=\partial_\mu  +[\Gamma_\mu,\ ],\
\Gamma_\mu={1\over 2}(u^\dagger\partial_\mu u+u\partial_\mu
u^\dagger),\
u_\mu=i(u^\dagger\partial_\mu u-u\partial_\mu
u^\dagger)\cr}\eqno(2)
$$

It was shown in Ref. [12] that the low-energy constants
$L_1,L_2,L_3$ (from which the Skyrme term derives) and $L_9$ (related to the
pion e.m. radius) are saturated to a very good approximation by the
resonance contribution from Eq. (1)

 Another important feature of the theory that Eq. (1)
defines, is its close resemblance with the
$\omega$-stabilized model. It is easy to observe that the canonical momentum
conjugate to the space components of the antisymmetric tensor field
vanishes identically: $\displaystyle{{\cal{\pi}}_{ij}=
{{\partial\Lag}\over{\partial[\partial_0 W^{ij}]}}=0}$. This
means that the components $\displaystyle{W_{ij}}$ of the
resonance tensor field are constrained. To gain more
insight into the problem, we go through the algorithm of constrained
dynamics [16], and find that the conservation in time of the primary
constraint leads  to:
$$\eqalign{
W_{ij}=&{1\over {2M_\rho^2}}\bigg\{-i\sqrt{2}G_\rho[u_i
,u_j]+\nabla_i {\cal{\pi}}_{0j}-\nabla_j {\cal{\pi}}_{0i}
\bigg\}\cr}\eqno(3)
$$

\noindent These
frozen fields are the ones which couple to the static soliton field.
 Anticipating on the next section, let us only observe that because of Eq.
(3), $\displaystyle{W_{ij}}$ cannot be trivially set to zero,
as a naive variational study of Eq. (1) could suggest.
This is exactly what happens in the $\omega$-stabilized case [10]. The
only difference is that in the present model one would encounter the
repulsion of the Skyrme term, while in the $\omega$ case it is a sixth
order term which is doing the job.

\noindent{\bfmagd III. Soliton solutions}

In terms of true degrees of freedom, the secondary Hamiltonian
functional is a function of  $\displaystyle{\vec F}$
and its canonical momentum $\displaystyle{\vec\phi}$ describing the pion
field,  and  of $\displaystyle{W_{0i}}$,  $\displaystyle{\pi_{0i}}$
representing the three polarizations of the massive spin-1
$\displaystyle{\rho}$ field and their respective momenta. We assume the
hedgehog ansatz for the pion, and the most general spherical ansatz for
the resonance fields:

$$\eqalign{
&\vec F=\hat r F,\quad W_{0i}=f_\pi\big[w_1(\tau_i-(\vec\tau.\hat r)\hat
r_i)+w_2(\vec\tau.\hat r)\hat r_i
-w_3(\vec\tau\times\hat r)_i\big]\cr
&\vec \phi=\hat r {{f_\pi^2}\over r}\phi,\quad \pi_{0i}={{f_\pi}\over
r}\big[\pi_1(\tau_i-(\vec\tau.\hat r)
\hat r_i)+\pi_2(\vec\tau.\hat r)\hat r_i
-\pi_3(\vec\tau\times\hat r)_i\big]
\cr}\eqno(4)
$$
\noindent  At the classical level, the radial functions
$\displaystyle{F,\phi,w_1,w_2,w_3,\pi_1,\pi_2,\pi_3}$ are
found by extremizing the Hamilton functional
which can be written as a sum of two terms:

$$\eqalign{
H=4\pi f_\pi^2\intif
dr\big[{\cal{H}}_Q(F,\phi,w_1,w_2,w_3,\pi_1,\pi_2)+{\cal{H}}_{\pi\rho}(F,\pi_3)
\big]
\cr}\eqno(5)
$$

\noindent The first piece, $\displaystyle{{\cal{H}}_Q}$,
is positive and {\it quadratic} in
the fields $\displaystyle{\phi,w_1,w_2,w_3,\pi_1,\pi_2}$:

$$\eqalign{
{\cal{H}}_Q=&{1\over{2}}(\phi+2g_\rho
w_3 \sin F)^2+{{\pi_2^2}\over 4}+{{\pi_1^2}\over 2}
+(r \dot w_2 +2w_2-2\cos F w_1)^2\cr
&+(M_\rho r)^2(w_2^2+2w_1^2+2w_3^2)
+{1\over{2M_\rho^2}}(\dot\pi_1-\cos F{{\pi_2}\over r})^2
\cr}\eqno(6)
$$

\noindent with  $g_\rho=2\sqrt{2}G_\rho / f_\pi$. As long as the fields are
nonsingular, it is straightforward to show that
$\displaystyle{{\cal{H}}_Q}$ vanishes in the static limit.
We stress that we do not impose
the $\displaystyle{\phi,w_1,w_2,w_3,\pi_1,\pi_2}$ components to be
identically zero. We rather find, by \ solving \ the \ static \ Hamilton
\ equations, that \ the \ extremum
of \ $\displaystyle{{\cal{H}}_Q}$ is met when
$\displaystyle{\phi=w_1=w_2=}$
$\displaystyle{w_3=\pi_1=\pi_2=0}$. In other words, the
possibility of having spontaneous parity violating (SPV) solutions which
exist in the Yang-Mills context (in the presence of stabilizing terms
 [9]), is ruled out here.

The second term in  Eq. (5) is:
$$\eqalign{
{\cal{H}}_{\pi\rho}=&{1\over 2}\big[(r\dot F)^2+2\sin^2
F\big]
+{{\pi_3^2}\over 2}\cr
&+{1\over{2M_{\rho}^2}}\big\{(g_\rho\sin F\dot F-\dot\pi_3)^2
+2(g_\rho{{\sin^2 F}\over {2r}}-\cos F{{\pi_3}\over
r})^2\big\}
\cr}\eqno(7)
$$
\noindent  We now look for static
soliton solutions carrying one unit of winding number i.e., satisfying
the boundary conditions $\displaystyle{F(0)=\pi}$
, $\displaystyle{F(\infty)=0}$ and require
$\displaystyle{{\cal{H}}_{\pi\rho}}$ to be stationary under
arbitrary variations of $\displaystyle{F}$ and $\displaystyle{\pi_3}$.
The set of (two) Hamilton equations that one obtains, can be solved only
numerically.  We find that  there is a solution for the whole range of
values of
$\displaystyle{g_\rho}$. In Fig. 1 we plot such a solution, for
$\displaystyle{M_\rho=769}$ MeV, $\displaystyle{f_\pi}$=93 MeV and
$\displaystyle{g_\rho=2.1}$ (from the $\rho$ width). The
classical soliton mass we find is $\displaystyle{M=1140}$ MeV, which is
slightly lower than the corresponding Skyrmion mass (1220 MeV) that one
finds when the $\rho$ mass and the magnitude of the coupling to the pion
are increased to infinity, keeping their ratio finite ($e=\sqrt{2}M_\rho /
f_\pi g_\rho$). At the classical level the properties of our solution
are very close to those of the saddle-point solution
of Ref. [11] where the hidden symmetry Lagrangian is used.

\noindent{\bfmagd IV. Classical stability}

The Hamiltonian (Eqs. (6) and (7)) is expressed in terms of {\it
independent} \rm field variables. One should perform arbitrary fluctuations
of these, and check whether the classical solution corresponds to
a local minimum. In the Yang-Mills approach the classical solutions are
destabilized by the fluctuations of positive parity [17]. In our
notations these correspond to the components $\displaystyle{w_1,w_2}$
(and $\pi_1,\pi_2$) which occur only in  $\displaystyle{{\cal{H}}_Q}$.
If one expands $\displaystyle{{\cal{H}}_Q}$  to quadratic order, one
easily sees
that the fluctuations of $w_1,w_2,\pi_1$ and $\pi_2$ as well
as those of $\phi$ and $w_3$ decouple from the fluctuations of $F$
and $\pi_3$ which occur only in ${\cal{H}}_{\pi\rho}$. Thus, the
contribution from $\displaystyle{{\cal{H}}_Q}$ is obviously
positive and does not induce any instability.

\noindent Let us now turn to $\displaystyle{{\cal{H}}_{\pi\rho}}$ (Eq. (7)).
As a preliminary stability check we performed the following global scaling
transformations of the Derrick [5] type: $\displaystyle{F(r)\to F(\lambda r)}$
 and $\displaystyle{\pi_3(r)\to\gamma\pi_3(\lambda r)}$. We
checked numerically that the matrix
$\displaystyle{\delta_{\lambda\gamma}^2{\cal{H}}_{\lambda\gamma}}$  has
two positive eigenvalues, for any value of $\displaystyle{g_{\rho}}$.

\noindent Consider now the more general
fluctuations  $\displaystyle{F=F_0(r)+\delta  F(r,t)}$,
$\displaystyle{\pi_3=\pi_3^0(r)+\delta \pi_3(r,t)}$
 (breathing fluctuations), which are the most dangerous for
the stability (see e.g. Ref. [19]). The variation of the Hamiltonian  up to
second order is :
$$\eqalign{
\delta{\cal{H}}_{\pi\rho}=2\pi f_\pi^2\intif \psi^\dagger{\cal{M}}\psi dr\cr}
\eqno(8)
$$
\noindent where $\psi$ is a two component vector with $\displaystyle{
\psi^\dagger=(\delta F \ \ \delta \pi_3)}$ and ${\cal{M}}$ is a hermitian
matrix operator of the radial variable $r$, which reads :
$$\eqalign{
&{\cal{M}}_{(\delta F,\delta F)}=-{d\over
{dr}}(r^2+{{g_\rho^2}\over {M_\rho^2}}\sin^2F_0){d\over {dr}}
+2\cos 2F_0\cr
&-{{g_\rho^2}\over {M_\rho^2}}\big[
\sin 2F_0 \ddot F_0+ \cos 2F_0 \dot F_0^2-{{\sin^2 F_0}\over
{r^2}}(3-4\sin^2 F_0)\big]\cr
&+{1\over{M_\rho^2 r^2}}\big[g_\rho \cos F_0(\ddot \pi_3^0 r^2+\pi_3^0
(9\sin^2 F_0-2))-2(\pi_3^0)^2\cos 2F_0\big]\cr
&{\cal{M}}_{(\delta \pi_3,\delta \pi_3)}={1\over
{M_\rho^2}}\big[-{{d^2}\over{dr^2}}+(M_\rho^2+2{{\cos^2 F_0}\over
{r^2}})\big]
\cr
&{\cal{M}}_{(\delta F,\delta \pi_3)}={1\over {M_\rho^2}}
\big[
g_\rho\sin F_0 {{d^2}\over {dr^2}}
+{1\over{r^2}}(-2\sin 2F_0\pi_3^0+g_\rho(3\sin^2 F_0-2)\sin F_0)
\big]
\cr
&{\cal{M}}_{(\delta \pi_3,\delta F)}=({\cal{M}}_{(\delta F,\delta
\pi_3)})^\dagger
\cr}\eqno(9)
$$
\noindent Observe that the first two lines in Eq. (9) are nothing but
the fluctuation operator of the Skyrmion [18]: this is because the
$\pi_3^0$ terms as well as the coupling potential generate terms of
order $1/M_\rho^4$ or higher.

To prove the stability, one has to show that $\displaystyle{{\cal{M}}}$
has no negative eigenvalues. We verified this using two different
methods

i) diagonalization of a discretized approximation of $\displaystyle{{\cal{M}}}$
: we found only positive eigenvalues

ii) evaluation of the Jost determinant for negatives values of the
energy: we found no sign change.

\noindent{\bfmagd V. Vector mesons in terms of vector fields}

Finally, one might raise the question of why is it necessary to resort
to antisymmetric tensor fields instead of just vector fields with
exactly the same chiral transformation law. To answer that question, let
us consider the following Lagrangian of vector fields [15]:

$$\eqalign{
{\cal{L}}_V=&-{1\over 4}\tr(V_{\mu\nu}V^{\mu\nu})+{{M_{\rho}^2}\over
2}\tr(V_\mu V^\mu)\cr
&+{{f_\pi^2}\over 4}\tr(\partial_\mu U\partial^\mu U^\dagger)
-i{{g_V}\over{2\sqrt{2}}}\tr(V_{\mu\nu} [u^\mu, u^\nu])\cr}\eqno(10)
$$

\noindent with $\displaystyle{V_{\mu\nu}=\nabla_\mu V_\nu-\nabla_\nu
V_\mu}$. This Lagrangian differs from the massive Yang-Mills one by $O(V^2)$
terms. Furthermore, vector meson exchange from Eq. (10) can only
produce terms starting at $O(p^6)$ in the chiral expansion.
In fact, this theory necessitates
{\it explicit} inclusion of the Skyrme term otherwise the Hamiltonian is
not positive. Indeed, one finds the following contribution to
${\cal{H}}_Q$ (in the hedgehog ansatz, denoting the radial functions
$v_i$ and $\bar\pi_i$):

$$\eqalign{
{\cal{H}}_Q={1\over 2}
{
{(\phi+2\sqrt{2}g_V\bar\pi_3\sin F)^2}
\over
{1-8g_V^2\sin^2 F/ f_\pi^2 r^2} }
+\  ... \cr}\eqno(11)
$$

\noindent  The Hamiltonian has a pole and is
{\it not bounded from below}. This is clearly unacceptable. The simplest
way to remove the pole
is to add to ${\cal{L}}_V$ a local term:

$$\eqalign{
{\cal{L}}_V'={\cal{L}}_V+{{g_V^2}\over 8}\tr([u^\mu, u^\nu][u_\mu,
u_\nu])\cr}\eqno(12)
$$

\noindent Furthermore, the Hamiltonian resulting from this operation is
equivalent to the Hamiltonian of the antisymmetric tensor field
approach. In fact, there is a {\it canonical} transformation between
the two sets  {$\vec F,\vec\phi,V_i,\bar\pi_i$}
and {$\vec F,\vec\phi,W_{0i},\pi_{0i}$} with $g_V=G_\rho/M_\rho$:

$$\eqalign{
V_i\to -{{\pi_{0i}}\over
{2M_\rho}},
\ \bar\pi_i\to 2M_\rho W_{0i}\cr}\eqno(13)
$$
\noindent An alternative motivation for replacing ${\cal{L}}_V$ by
${\cal{L}}_V'$ is provided in Ref. [15].

\leftline{\bfmagd Aknowledgements}
We are grateful to Profs. R. Vinh Mau and R. Kerner for critical comments
and discussions.
\bigskip
\leftline{\bfmagd References}
\bigskip
\item{[1]}T. H. R. Skyrme, Proc. Roy. Soc. A260 (1961) 127
\medskip
\item{[2]}G. 't Hooft, Nucl. Phys. B79 (1974) 276
\medskip
\item{[3]}E. Witten, Nucl. Phys. B179 (1979) 57
\medskip
\item{[4]}G. Adkins, C. Nappi and E. Witten, Nucl. Phys. B228 (1983) 552
\medskip
\item{[5]}G. H. Derrick, J. Math. Phys. Vol.5  No9 (1964) 1252
\medskip
\item{[6]}K. Iketani, Kyushu Univ. preprint, 84-HE-2 (1984)
\medskip
\item{[7]}U. G. Meissner and I. Zahed, Phys. Rev. Lett. 56 (1986) 1035;
Z. Phys. A327 (1987) 5 ; M. Lacombe, B. Loiseau, R. Vinh Mau and
W. N. Cottingham,  Phys. Rev. Lett. 57 (1986) 170 ;
Phys. Rev. D 38 (1988) 1491; M. Chemtob, Nucl. Phys. A 466 (1987) 509
\medskip
\item{[8]}G. Adkins and C. Nappi, Phys. Lett. B137 (1984) 552
\medskip
\item{[9]}A. Kobayashi, H. Otsu, T. Sato and S. Sawada, Nagoya Univ.
preprint, DPNU 91-50, IPC-91-04 ; K. Yang, S. Sawada, A. Kobayashi, Prog.
Theor. Phys. 87 (1992) 457
\medskip
\item{[10]}H. Forkel, A. D. Jackson and C. Weiss, Nucl. Phys. A526 (1991) 453
\medskip
\item{[11]}Y. Igarashi, M. Johmura, A. Kobayashi, H. Otsu, T. Sato and S.
Sawada, Nucl. Phys. B259 (1985) 721.
\medskip
\item{[12]}G. Ecker, J. Gasser, A. Pich and E. de Rafael, Nucl. Phys.
B321 (1989) 311
\medskip
\item{[13]}J. Gasser and H. Leutwyler, Ann. Phys. (N.Y.) 158 (1984) 142
\medskip
\item{[14]}S. Coleman, J. Wess and B. Zumino, Phys. Rev. 177 (1969) 2239
; C. G. Callan, S. Coleman, J. Wess and B. Zumino, {\it ibid} p. 2247
\medskip
\item{[15]}G. Ecker, J. Gasser, H. Leutwyler, A. Pich and E. de Rafael
, Phys. Lett.  B223 (1989) 425
\medskip
\item{[16]}K. Sundermayer, {\it Constrained Dynamics} in Lecture Notes
in Physics, Springer-Verlag 1982
\medskip
\item{[17]}Z. F. Ezawa and T. Yanagida, Phys. Rev. D33 (1986) 247 ;
J. Kunz and \hfill\break D. Masak , Phys. Lett. B179 (1986) 146
\medskip
\item{[18]}I. Zahed, U. G.  Meissner and  U. B. Kaulfuss, Nucl. Phys.
A426 (1984) 525 ; \hfill\break A. Abada and D. Vautherin , Phys. Rev. D46
(1992)
\medskip
\item{[19]}Y. Brihaye, C. Semay and J. Kunz, Phys. Rev. D44 (1991) 250
\bigskip
\eject
\leftline{\bfmagd Figure caption}
The solution extremizing the Hamiltonian functional of Eq. (7). The
full line displays the chiral
function $F(r)$ in the present model, while the dotted one is the solution
of the corresponding Skyrme model. In the dashed-dotted line we display
$-\pi_3(r)$, the $\rho$ degree of freedom.
\eject

\bye